\documentstyle[12pt,fleqn]{article}

 \csname
@addtoreset\endcsname{equation}{section}

\title{Explosive crystallization mechanism\\ of ultradisperse
amorphous films} \author{ A.~I.~Olemskoi$^{*}$,
A.~V.~Khomenko$^{*}$, V.~P.~Koverda$^{**}$ \\ {\it $^{*}$Sumy State
University, Rimskii-Korsakov St., 2,} \\ {\it 244007 Sumy, Ukraine,
E-mail:  olemskoi@ssu.sumy.ua }\\ {\it $^{**}$ Institute of Thermal
Physics,}\\ {\it Urals Branch of Russian Academy of Sciences} \\ {\it
620219 Ekaterinburg, Russia, E-mail:  iva@itp.e-burg.su} } \date{\ }

\begin{document}
\maketitle{
\vspace*{-0.8cm}

\thispagestyle{empty}
\noindent {\bf Abstract}
\vspace{0.3cm}

The explosive crystallization of germanium ultradisperse amorphous
films is studied experimentally. We show that
crystallization may be initiated by local heating at the
small film thickness but it realizes spontaneously at the large
ones. The fractal pattern of the crystallized phase is
discovered that is inherent in the phenomena of diffusion limited
aggregation. It is shown that in contrast to the ordinary
crystallization mode the explosive one is connected with the
instability which is caused by the self-heating.
A transition from the first mechanism to the second one
is modelled by Lorenz system.
The process of explosive crystallization  is represented on the basis
of the self-organized criticality conception. The front
movement is described as the effective diffusion in the ultrametric
space of hierarchically subordinated avalanches, corresponding to the
explosive crystallization of elementary volumes of ultradisperse
powder. The expressions for the stationary crystallization heat
distribution and the steady-state heat current are obtained. The heat
needed for initiation of the explosive crystallization is obtained as
a function of the thermometric conductivity. The time dependence of
the spontaneous crystallization probability in a thin films is
examined.

\vspace{0.4cm}

\noindent PACS number(s): 81.30.-t, 05.40.+j, 64.60.Lx

\noindent {\it Keywords}: explosive crystallization; self-organized
criticality; ultrametric space


}


\section{Introduction}\label{sec1}

The metastable amorphous films are obtained usually by quenching of
a melt or by steam condensation  on a cold substrate \cite{1}--\cite{4}.
Experiments show a vast variety and complicated character
of the subsequent transitions into the stable crystalline state
\cite{1}--\cite{8d}. If the film thickness is so small
that the crystallization heat can be absorbed by a thermostat,
the ordinary cold crystallization mechanism works \cite{1}.
So, in the crystallized films of semiconductors
the undulating surface is developed under the formation of combs,
whose long axes are perpendicular to the direction
of the crystallization front movement \cite{7d}.
As is known \cite{7d,8d}, at small number of the crystal embryos
this mechanism is realized if
the nucleation frequency of crystallization centers $J$ is very small.

Sometimes, explosive crystallization can be initiated
by the local heating (for example, by laser or electron impulse).
Such a scenario takes place in the case of instability appearance
of the interfacial boundary motion due to both the heat exchange
with substrate and the influence of laser radiation \cite{7d,8d}.
This instability is ensured with a nonlinear dependence
of the crystal growth velocity $u$ as a function of temperature.
Moreover, the crystallization front instability can be
fluctuational in character that manifests itself in the
experiments with undulation of crystallized surface \cite{7d,8d}.
Such a behaviour appears in the partially crystallized film or at the
incomplete crystallization of amorphous phase.

Another scenario is observed in the amorphous medium in which as the
crystal growth velocity $u$, so the nucleation rate $J$ of embryonic
crystals are sufficient large. Indeed, at low temperatures, the
quantities $J$ and $u$ increase with temperature growth, so that the
self-heating stimulates the crystallization.  Therefore, the increase
of a film thickness can lead to a situation, when the crystallization
heat can not be absorbed by an environment that causes the heat
instability \cite{5}.  As a result, the spontaneous transition to
regime of the explosive crystallization can be provided by the
heating effect.  Examples of such amorphous mediums are the amorphous
ice layers, the some organic matters \cite{1,3}, and the layers of
the germanium amorphous ultradisperse powder with the admixture of
the crystalline phase \cite{3,4}.

Our work is devoted to studying of the explosive crystallization
mechanism that is provided by such type instability.

The experimental data in Sec.~\ref{sec2} show that course
of explosive crystallization of ultradisperse amorphous materials is
determined by high density of crystalline phase embryos. The
crystallization phase spreading is similar to percolation cluster
formation under consideration of liquid flow in a
random medium \cite{6}.  The formed cluster has the branching fractal
structure that is characteristic for the thermal conductivity limited
aggregation. Our approach is based on the assumption that such
structure leads to the hierarchical picture of the explosive
crystallization process.

In Sec.~\ref{sec3} investigation of realization conditions of this
picture  as a result of self-organization is carried out.  This
process is fixed by the velocity of crystallization front motion, its
temperature, and difference of the thermodynamic potentials of
amorphous and crystalline states. The first of mentioned parameters
is connected with the third one by means of the positive feedback
that is the reason for self-organization. The connection between the
first and second parameters is due to the negative feedback
reflecting the Le Chatelier principle.  As a result of the interplay
between the pointed out factors, the stationary state is established
at supercritical value of transformation thermal energy, where the
crystallization front velocity can take anomalous large values.
Then, at small film thickness, a regime is realized when the
crystallization process can be initiated only by the external
influence type of the laser beam \cite{2}.  However, with film growth
up to the critical value the crystallization heat in the film volume
becomes to be sufficient for spontaneous increase of the front
velocity.  Such a situation is observed experimentally in Ref. \cite{3}.

In Sec.~\ref{sec4} the direct examination of explosive
crystallization is carried out as a self-organized
criticality process caused by stochastic heat spreading over
hierarchical tree nodes. The study of effective motion equation
shows that, in accordance with the result of previous section,
the instability is developing in the case, when crystallization heat
effect (or externally inputted energy) is above a critical
quantity determined by the thermometric conductivity.
The stationary crystallization heat distribution, defined as a
solution of corresponding Fokker-Planck equation, allows us to
find both the heat current arising during crystallization and the
probability of spontaneous crystallization in a film with subcritical
thickness.  According to Sec.~\ref{sec5}, the probability increases
logarithmically as a function of time until the maximum value.  On
its turn, this value decreases monotonically with the growth of
thermometric conductivity.

\section{Experimental results}\label{sec2}

The experimental study of influence of the crystal inclusions distributed in
the amorphous phase volume in the kinetics of the explosive
crystallization has been carried out with the germanium \cite{4}.
Unfortunately, there are
no information in literature about spontaneous rise of the explosive
crystallization in continuous amorphous thin films of germanium.
It is known only that the maximum value
of formation frequency of crystalline embryos is
$J_{max} \sim 10^{14}$s$^{-1}$cm$^{-3}$ in the supercooled germanium
 and corresponds to the more high temperatures than
realized at the explosive crystallization process \cite{1}.
Therefore, in the germanium amorphous films the
natural process of crystals nucleation  has not enough intensity
for significant influence in the explosive crystal growth.
Let us point out in this connection that in amorphous ice layers, where
the spontaneous explosive crystallization takes place, one has
$J_{max} \sim 10^{20}$s$^{-1}$cm$^{-3}$.

In order to intensify the mentioned influence of crystal inclusions,
 the experiments were carried out with the thin films of amorphous
nanopowders with admixture of large number of smallest crystals
 having not more 3--10\% of total mass. The layers of amorphous
powder with the characteristic size of particles  3--10 nm were
obtained by thermal evaporation and following condensation of
germanium in the atmosphere of inert gas at the pressures 10--100 Pa.
Changing the evaporation intensity allows us to regulate the part of
the crystalline particles in amorphous powder.  Another peculiarity
of our experiments is that the substrates absorb significantly
smaller heat due to porosity of amorphous films.

The spontaneous explosive crystallization has been observed in layers
of nanopowders with the thickness 0.01--0.1 mm at the substrate
temperature 300--400 K.  In dependence on initial concentration of
crystalline phase, the movement velocities of crystallization wave
have been changed in the range 0.01--0.1 m/s.  In contrast to the
transverse undulation being inherent in the usual crystallization
mechanism, in our case the front movement leads  almost always to
formation of "twigs" along the movement direction.  The
characteristic pictures of explosive crystallization in the powder
layers is shown in Fig.~1, where the light background corresponds to
non-crystallized domains.  It is seen that the rise of explosive
crystallization avalanches occurs from the single centers which act
as original embryos.  The cornerstone of our observation is that the
crystalline phase distribution has a fractal character being similar
to the pattern appeared in diffusion limited aggregation \cite{6}.

\section{Determination of explosive crystallization conditions}\label{sec3}

The experimental data show that two mechanism of amorphous material
crystallization, depending on external conditions and presence of
crystalline phase embryos, are possible: the slow growth of a cold
crystal and the explosive crystallization that is caused by the
phase transformation heating. According to \cite{4}, the
transition between these regimes is jump-like in character, as a first-order
phase transition. We will show below that such a transition is caused by
the system self-organization due to the positive feedback between
the heating and growth velocity of crystalline phase.

To analyze the problem, let us examine the time dependencies of the
crystallization front velocity $u(t)$, its temperature
$T(t)$, and the specific crystallization heat $f(t)$. The equations
defining these dependencies take into account their
dissipative character and the positive feedback between quantities
$u$ and $f$, that is the reason for self-organization.  On the other
hand, in order to provide the stability of a system we introduce also
the negative feedback between $u$ and $T$. The obtained as a result
equations coincide formally with the Lorenz system that is the
simplest way to describe the self-organization process \cite{7}.

The first of the stated equations has the form
\begin{equation} \dot u = - u /\tau_u + \mu T,  \label{1} \end{equation}
where the dot stands for a derivative with respect to  time $t$,
$\mu>0$ is a constant. The first term in the right-hand side describes
the Debye relaxation during time $\tau_u$, the second one reflects
the increase of the  crystallization front velocity with the growth of
temperature difference $T=T_0-T_\infty$ at the crystallization front
and the thermostat, respectively. In the stationary state $\dot u =
0$ and Eq.~(\ref{1}) gives linear relationship $u=A_uT$, where the
constant $A_u \equiv \tau_u \mu$ is introduced.

The equation for the rate of quantity $T$ variation has the nonlinear
form
\begin{equation} \dot T = - T /\tau_T + g_T u f, \label{3}
\end{equation} where $f>0$ is the volume density of difference of the
thermodynamic potentials of amorphous and crystalline states;
$\tau_T,~g_T$ are positive constants. As in Eq.~(\ref{1}), the
first term in the right-hand side of Eq.~(\ref{3}) describes the
relaxation process of temperatures difference $T$ to the stationary
value $T = 0$. It takes place not during the macroscopic time
$\tau_u$ but during the mesoscopic one $\tau_T$, so that the
important for the future condition $\tau_T \ll \tau_u$ is satisfied.
The second term describes the mentioned positive feedback between the
crystallization front velocity $u$ and the difference $f$ between the
specific thermodynamic potentials of the phases that results in the increase
of value $T$ and, thus, causes the self-organization process. In the
stationary case $\dot T = 0$ Eq.~(\ref{3}) takes form
\begin{equation} T_0 = T_\infty + A_T u f, \label{4} \end{equation}
where the constant $A_T \equiv g_T \tau_T$ is introduced. According
to Eq.~(\ref{4}) nonlinear term describes the heating of the
crystallization front with the growth of crystallization wave
velocity.

The kinetic equation for the difference $f$ of specific thermodynamic
potentials
\begin{equation} \dot f = (f_0-f)/\tau_f - g_f u T \label{6}
\end{equation} differs from Eqs.~(\ref{1}), (\ref{3}) as follows:
the relaxation of quantity $f$ occurs not to the zero but to the
finite value $f_0$, representing the energy density inputted in the
system (heat effect of transformation); $\tau_f$ is
a corresponding relaxation time. In Eq.~(\ref{6}) the negative feedback
between the quantities $u$ and $T$ is introduced to imply the decrease
of thermodynamic transformation factor $f$ with the growth of the
crystallization front velocity and its temperature ($g_f>0$ is
a corresponding constant).

Let us study the system of differential equations
(\ref{1}), (\ref{3}), (\ref{6}) defining the self-consistent behaviour
of the quantities $u(t), T(t), f(t)$ which act as the order
parameter, the conjugate field, and the control parameter,
respectively \cite{7}.
With this aim, we write the Lorenz system in the form
\begin{eqnarray}
&&\tau_u\dot u = -u + A_uT, \nonumber \\
&&\tau_T\dot T = - T + A_T u f, \label{8} \\
&&\tau_f\dot f = (f_0-f) - A_f u T,  \nonumber
\end{eqnarray}
where the constant $A_f \equiv \tau_f g_f$ is introduced.
As is known \cite{7}, the analytic examination is possible only provided that
hierarchical subordination conditions are satisfied \begin{equation}
\tau_T\ll\tau_u, \quad \tau_f\ll\tau_u, \label{7} \end{equation}
which mean that in the course of evolution the temperatures
difference $T$ and the thermodynamic potential $f$ follow the
variation of the crystallization front velocity.
As was mentioned above, the first
of these conditions is always obeyed. Since, on the other hand,
$\tau_f\sim\tau_T$ the second inequality (\ref{7}) is met also.

When the values of relaxation times
$\tau_u,~\tau_T,~\tau_f$ are constant and the conditions (\ref{7})
are obeyed, it is not difficult to see that
the system of equations (\ref{8}) describes the
second-order transition. However, the cold crystallization mode
transforms into the explosive one in accordance with the first-order
mechanism.  To avoid this discrepancy, let us use the
simplest approximation \cite{8} \begin{equation} {1 \over \tau_u} =
{1 \over \tau_0}\left( 1 + {\kappa \over 1+\left(u/u_{\tau} \right)^2
} \right), \label{12} \end{equation} characterized by the positive
constants $\tau_0, \kappa$, and $u_{\tau}$.
Moreover, it is convenient to introduce the
scales of quantities $u, T$, and $f$:  \begin{equation} u_m\equiv
(A_TA_f)^{-1/2}, \quad T_m\equiv u_m/A_u=A_u^{-1}(A_TA_f)^{-1/2},
\quad f_c\equiv (A_uA_T)^{-1}, \label{13} \end{equation} where
$A_u\equiv \tau_0 \mu$. Then, the system (\ref{8}) assumes the
simplest form
\begin{eqnarray}
&&\tau_0\dot u = - u\left[ 1 + \kappa
(1+ u^2/\alpha^{2})^{-1} \right] + T, \label{14} \\
&&\tau_T\dot T = - T + u f, \label{15}\\
&&\tau_f\dot f = (f_0-f) - u T, \label{16}
\end{eqnarray}
where the parameter $\alpha\equiv u_\tau/u_m$ is introduced.

Taking into consideration conditions (\ref{7}), the left-hand sides of
Eqs.~(\ref{15}), (\ref{16}) can be set equal zero because of
the small relaxation times $\tau_T,~\tau_f$.  As a result we
obtain the expressions for the temperatures difference $T$ and the
thermodynamic potential $f$ in terms of the velocity $u$ of
crystallization front:  \begin{eqnarray} T&=&f_0 u (1+u^2)^{-1},
\label{16d} \\ f&=& f_0 \left( 1+ u^2 \right)^{-1}.  \label{17d}
\end{eqnarray} At $u\ll 1$, the dependence (\ref{16d}) has the
linear form characterized by susceptibility $\partial u /\partial
T = f_0^{-1}$.  At $u=1$ function $T(u)$ becomes saturated, and at
$u>1$ decreasing dependence is realized  that has no physical
meaning. This implies that the constant $u_m$ defined in (\ref{13})
has the physical meaning of the maximum value of the crystallization front
velocity. According to Eq.~(\ref{17d}) the difference $f$ of specific
thermodynamic potentials of different phases decreases monotonically with the
growth of velocity $u$ from the value $f_0$ at $u=0$ to the $f_0/2$
at $u=1$. Obviously, this decrease is caused by the negative
feedback in Eq.~(\ref{16}) that is the reflection of Le Chatelier
principle for examined problem. On the other hand, the positive feedback
between the velocity $u$ and the thermodynamic factor $f$ in
Eq.~(\ref{15}) is the reason for transition to the explosive
crystallization mechanism that leads to the growth of $T$ due to
crystallization front heating. However, in accordance with Eq.~(\ref{16}),
the latter results in decrease of $f$ to be
a consequence of self-organization process.

Within the framework of the adiabatic approximation $\tau_T, \tau_f
\ll \tau_0$, the Lorenz system (\ref{14})--(\ref{16}) is reduced to the
Landau-Khalatnikov equation
\begin{equation} \tau_0 \dot u = - \partial V/ \partial u.
\label{18} \end{equation} Its form is determined by the effective
potential \begin{equation}
V{=} {1\over 2} \left[ u^2 {-} \theta \ln \left( 1{+}u^2
\right)\right]{+} {\kappa \alpha^2\over 2} \ln \left[
1{+}\left({u/\alpha}\right)^2 \right], \label{19} \end{equation}
where $\theta \equiv f_0/\Delta h,~ \Delta h \equiv
\left(\tau_0\tau_T g_u g_T\right)^{-1}$ is the scale defining
specific crystallization heat, the quantity $V$ is measured in
units of $u_m^2$. For small values of $\theta$ the curve of the $V$ vs $u$
dependence has a monotonically increasing shape with its minimum at
point $u = 0$ that corresponds to the cold crystallization mechanism.
At $\theta=\theta_c^0$, where
\begin{equation} \theta_c^0 \equiv 1 + \alpha^2 (\kappa-1) + 2 \alpha
\sqrt {\kappa \left( 1 - \alpha^2\right)} \label{20} \end{equation}
a plateau appears, which for $\theta > \theta_c^0$ is transformed
into a minimum corresponding to the velocity $u_e \not= 0$ and a
maximum at point $u^m$ separating a minima which meet to the
values $u = 0$ and $u = u_e$. When the parameter $\theta$ increases
still further, the minimum at point $u=u_e$ is lowered and the
height of barrier at $u=u^m$ decreases, vanishing at the critical value
\begin{equation} \theta_c = 1+\kappa. \label{21} \end{equation}
The steady-state values of the crystallization front velocity have
the form
\begin{eqnarray}
&&u_e^m = u_{00}
\left\{1\mp\left[ 1+ \left( {\alpha \over u_{00}^2}\right)^2 (\theta
- \theta_c) \right]^{1/2} \right\}^{1/2}, \label{22} \\
&& u_{00}^2 \equiv {1 \over 2} \left[ \left( \theta - 1 \right) -
\left( 1+\kappa \right) \alpha^2 \right]. \nonumber \end{eqnarray}
As is shown in Fig.~2a, if the system's energy increases slowly, the
jump from zero to $\sqrt{2}u_{00}$ is observed at point $\theta =
\theta_c$ and then the value $u_e$ increases smoothly. If the
parameter $\theta$ goes downward quasistatically, the crystallization
front velocity $u_e$ smoothly decreases up to the point, where
$\theta =\theta_c^0$ and $u_e=u_{00}$, and then jump-like
goes down to zero. The hysteresis of such type takes place
only at the presence of energy barrier inherent in a first-order
transition and appears if only the parameter $\alpha =
u_\tau / u_m$ is smaller than unity.

The key point of studied transition is that the
stationary value of the thermodynamic transition factor
\begin{equation} f_e = {\left( 1+u_{00}^2 \right)-\sqrt{ \left(
1+u_{00}^2 \right)^2 - \theta \left( 1-\alpha^2 \right)} \over
1-\alpha^2 } \label{24} \end{equation} equals the thermal energy
$\theta$ in the  $0< \theta < \theta_c^0$ interval
(Fig. 2b). At $\theta > \theta_c^0$ this quantity decreases
smoothly from the value
\begin{equation} f_{m} = 1 + \alpha[\kappa /
(1-\alpha^2)]^{1/2} \label{25} \end{equation} at $\theta =
\theta_c^0$ to 1 at $\theta \to \infty$.

Under quasistatic growth of parameter $\theta$ from 0 to $\theta_c$
the stationary value of transformation factor increases linearly
being in the same interval.  After jump-down at $\theta = \theta_c$
the quantity $f_e$ decreases smoothly according to dependence
(\ref{24}). Under reverse decrease of $\theta$ the quantity $f_e$
undergoes the jump at point $\theta_c^0$ from the value $f_m$ up to
the $\theta_c^0$. Since, in the important range of values of the
parameters $\alpha$ and $\kappa$ limited by $\kappa_{min} = \alpha^2
/ (1 - \alpha^2)$, the maximum value $f_m$ of specific transition
energy is smaller than the minimum value $\theta_c^0$ of the heat
density, the stationary value $f_e$ of the specific difference of
thermodynamic potentials of amorphous and crystalline states is
always smaller than heat density $\theta$.

The above analysis shows that  the effective potential $V(u)$ has the
barrier separating the cold and the explosive crystallization modes.
As heat density $\theta$ becomes greater than the critical
value $\theta_c$ this barrier disappears. Thus,  at
$\theta<\theta_c$ the transition to the explosive crystallization
mechanism requires the penetrating of energy barrier and at opposite
case it realizes spontaneously. The first of appointed situations
takes place in the case when the explosive
crystallization is initiated by an external beam (see Fig.~1a).
With the increase of the coating thickness the
crystallization heat can not be absorbed by substrate and
parameter $\theta$ increases.  This leads to that the value
$\theta_c$ (at which the function $V(u)$ loses barrier) is reached at
the critical film thickness and the system  transforms
into the explosive crystallization regime spontaneously (Fig.~1b).

\section{Description of explosive crystallization as a self-organized
criticality process} \label{sec4}

In recent years considerable study has been given to the conception
of self-organized criticality (SOC) representing the natural
development of critical phenomena picture \cite{9,10}. The basic
distinction of the SOC from the phase transition is that SOC process
realizes spontaneously, whereas the phase transition goes on under
the external influence (for instance, the temperature variation). In
this section we will expound the quantitative picture within the
framework of which the explosive crystallization will be represented
as SOC. The basis for the such representation is that the SOC process
consists of the hierarchically
subordinated sequence of elementary phase transitions which are
usually called avalanches \cite{11}. Their hierarchical subordination
manifests itself in that the avalanches of upper level are formed after
their formation has been finished on the lower one. Then this process
recurs on the more upper levels -- right up to the global avalanche
formation on the top of hierarchical tree. The
hierarchical nature of explosive crystallization process is
discovered obviously in the microscopic photographs of
crystallization pattern in Fig.~1, where it appeares
as the tree-like fractal structure.  In accordance with \cite{12}
the hierarchical tree represents the geometrical shape of ultrametric
space in that system's states are realized. Thus, for description of
explosive crystallization the geometrical picture of the nodes
distribution over hierarchical tree levels is necessary to represent,
at first \cite{13}.

Let maximum number of nodes $N$ be on the bottom
hierarchical level corresponding to the distance in the ultrametric
space $s=0$. This level meets the elementary avalanches whose number
coincides with $N$. There is the only node on the top level
($s=s_0\gg 1$) corresponding to the global avalanche. The problem is
to find the dependence $N(s)$ that define the distribution of tree
node number over hierarchical levels.

At first, we examine the basic types of the trees (Fig.~3): regular
tree with integer branching ratio $j$, regular Fibbonachi tree with
fractional $j=\tau\approx 1.618$, degenerate tree with the only
branching node per level and the tree of our primary concern
-- irregular tree. Let $k$ be the numbering index for the levels, so
that $k$ increases from the top level to the bottom one. The
variable
\begin{equation}
s=s_{0}-k \label{17a}
\end{equation}
then defines the distance in the ultrametric space \cite{12,14}.
Geometrically, objects of this space correspond to the nodes of the
bottom level ($k=s_{0}$) of a Cayley tree. Since the distance between
the nodes is defined by the number of steps to a common ancestor, the
distance is eventually the level number (\ref{17a}), counted from the
bottom.

As it can be seen in Fig.~3a, in the simplest case of regular tree
with integer branching ratio $j$ the number of avalanches
$N_{k}=j^{k}$ exponentially decays to zero with the distance
$s$ between them:
\begin{equation}
N(s)=N \exp{(-s\ln j)},\qquad N\equiv j^{s_{0}}.
\label{18a}
\end{equation}
In Eq.~(\ref{18a}) the equality (\ref{17a}) is used and the avalanche
number $N$ is related to the total number of levels $s_{0}$. For the
Fibbonachi tree (see Fig.~3b), where
$N_{k}=\nu\tau^{k},\,\nu\approx 1.171,\,\tau\approx 1.618$,
we have \begin{equation} N(s)=N \exp{(-s\ln \tau)},\qquad N\equiv
\nu\tau^{s_{0}}.  \label{19a} \end{equation} When Eq.~(\ref{19a}) is
 compared with Eq.~(\ref{18a}), it is clear that the exponential
decay remains unaltered in the case of fractional branching ratio and
characterizes the regularity of tree.

For the degenerate tree (see Fig.~3c) $N_{k}=(j-1)k+1$ and
Eq.~(\ref{17a}) provides the following linear dependence
\begin{equation}
N(s)=N-(j-1)s,\qquad N\equiv (j-1)s_{0}+1.
\label{20a}
\end{equation}
It can be shown that in the case of irregular tree, displayed in
Fig.~3d, the power law dependence is realized:
\begin{equation}
N_{k}=k^{a},\qquad a>1.
\label{21a}
\end{equation}
The latter can be regarded as an intermediate case between the
exponential Eqs.~(\ref{18a}), (\ref{19a}) and linear Eq.~(\ref{20a})
obtained for the limiting cases of regular and degenerate trees, respectively.
Formally, the approximation (\ref{21a}) means that a function $N(x)$
defined on the self-similar set of hierarchically subordinated
avalanches is homogeneous, $N(kx)=k^{a}N(x)$.
It is convenient to rewrite Eq.~(\ref{21a}) in
terms of the distance:
\begin{equation} N_{k}=N(1-s/s_{0})^{a},\qquad N\equiv
s_{0}^{a},\qquad a>1.
\label{22a}
\end{equation}

At given value of  crystallization thermal effect $Q_{k}$ the
heat current density between different levels $k$ is expressed by the
generalized Onsager equality
\begin{equation} j_{k}=-\chi(Q_{k})
\frac { {\rm d} Q_{k} } { {\rm d} k}. \label{23a} \end{equation}
Here, within the multiplicative noise representation
\cite{UFN} the effective thermometric conductivity coefficient
\begin{equation}
\chi(Q)=\chi Q^{\beta} \label{24a} \end{equation} is defined by the
constant $\chi>0$ and the exponent $\beta$.
The cornerstone of our approach is that
total heat current at given level is independent of the hierarchical
level:  \begin{equation} j_{k}N_{k}={\rm const}\equiv J.  \label{25a}
\end{equation} Substitution of Eqs.~(\ref{22a})--(\ref{24a}) into
Eq.~(\ref{25a}) gives the expression for the crystallization heat
effect: \begin{equation} Q_{k}=Qk^{-b},\qquad b=(a-1)/(1+\beta)>0
\label{26a} \end{equation} normalized by the maximum value $Q\equiv
Q_{k=1}$. Inserting Eq.~(\ref{17a}), we get the dependence
\begin{equation}
Q(s)=q(1-s/s_{0})^{-b}, \label{27a} \end{equation} where the heat
effect at the bottom level $s=0$ is \begin{equation} q\equiv
Qs_{0}^{-b}=QN^{-b/a}.  \label{28} \end{equation}

In general case, the condition of current conservation (\ref{25a})
is not satisfied and with accounting Eqs.~(\ref{26a}),
(\ref{28}) we assume the scaling relation
\begin{equation}
Q_{k}=N^{b/a}k^{-b}q_{k}, \label{29} \end{equation} where $q_{k}$ is
a slowly varying function to be determined. According to
Eqs.~(\ref{23a})--(\ref{25a})
it obeys the Landau-Khalatnikov equation
\begin{equation}
\frac{{\rm d}x}{{\rm d}\kappa}= -\frac{\partial V}{\partial x},
\label{30} \end{equation} where one denotes \begin{equation}
\kappa\equiv\ln{k^{b}},\qquad x\equiv q_{k}/q_{c},\qquad
q_{c}^{1+\beta}\equiv\left(J/b\chi\right)N^{-(a-1)/a},
\label{31}
\end{equation}
and the effective potential is
\begin{equation}
V=\frac{x^{1-\beta}}{1-\beta}-\frac{x^{2}}{2}. \label{32}
\end{equation}

As indicated in Fig.~4, this potential reaches its maximum value
$V_{0}=(1+\beta)/2(1-\beta)$ at $x=1$ and decreases indefinitely at
$x>1$. So, in order to initiate the process of explosive
crystallization, a low intensity avalanche with $q<q_c$ at the bottom
level needs to penetrate the barrier $V_{0}$. For study of
this process we proceed with stochastic Langevin equation with a
white noise (cf. Eq.~(\ref{30}))
\begin{equation}
\frac{{\rm d}x}{{\rm d}\kappa}= -\frac{\partial V}{\partial x}+\zeta,
\label{33} \end{equation} \begin{equation}
\langle\zeta\rangle=0,\qquad
\langle\zeta(\kappa)\zeta(\kappa^{\prime})\rangle=
2\chi\delta(\kappa-\kappa^{\prime}),
\label{34}
\end{equation} where the noise intensity $\chi$ is reduced to the
thermometric conductivity in Eq.~(\ref{25a}).

The solutions of this equation are distributed in the ultrametric
space according to the function $w(\kappa,x)$ that obeys the
Fokker-Planck equation \cite{15}:  \begin{equation} \frac{\partial
w}{\partial \kappa}+ \frac{\partial j}{\partial x}=0,\qquad
j\equiv-w\frac{\partial V}{\partial x}-\chi \frac{\partial
w}{\partial x}. \label{35} \end{equation} Since there is no current at
the equilibrium state ($j=0$), the distribution function
\begin{equation} w_{0}(x)\propto \exp{(-V(x)/\chi)} \label{36}
\end{equation} is dictated by the potential (\ref{32}). In the case
of non-equilibrium steady state the probability density $w$ does not
depend on the hierarchical level variable $\kappa$ and the current $j$
being constant, in compliance with conservation law (\ref{25a}), can
take a nonzero value. In accordance with Eq.~(\ref{35}) the
stationary $w(x)$ and the equilibrium $w_{0}(x)$ distributions are
connected by the equation \cite{16}:  \begin{equation}
\frac{w(q)}{w_{0}(q)}=\frac{j}{\chi}
\int\limits_{q/q_{c}}^{\infty}\frac{{\rm d}x}{w_{0}(x)}, \label{37}
\end{equation} where the boundary condition $w\to 0$ at $q\to\infty$
is taken into account.

Given the heat effect $q$ Eq.~(\ref{37}) allows the current $j$ to be
found. In trying to do it, special consideration should be given to
the fact that the heat $q$ is bounded from below, $q>G$ \cite{10}. The
appearance of the gap $G$ is the feature inherent in hierarchical
ensemble of crystallization centers. Indeed, after merging of them
within a hierarchical cluster of the size $s_{g}$, all $s$, such that
$s<s_{g}$, are appeared to be dropped out the consideration as well
as low heat effects with $q<q(s_{g})\equiv G$ (see Fig.3). The
expression for the current $j$ then can be derived from
Eq.~(\ref{37}) with the second boundary condition $w(G)=w_{0}(G)$.
The result reads
\begin{equation} j=2\chi W\left[ 1+{\rm erf}\left(
\sqrt{\frac{1+\beta}{2\chi}} \left(1-\frac{G}{q_{c}}\right)
\right)\right]^{-1}, \label{38} \end{equation} where the factor
\begin{equation}
W\propto\exp(-V_0/\chi), \qquad
V_0\equiv \frac{1}{2}~\frac{1+\beta}{1-\beta}
\label{39}
\end{equation}
gives the probability that fluctuation will surmount the barrier $V_0$ of
the potential (\ref{32}). Eq.~(\ref{38}) shows that in the case of
small gap, $G\ll q_{c}$, the current $j$ has the value $W\chi$ and it
is doubled under $G=q_{c}$. It can be understood if we picture the
effect of the gap as a mirror that reflects diffusing particles at
the point $q=G$: if $G\ll q_c$ a particle penetrating the barrier can
move along both directions, but in the case of $G=q_c$ the mirror is
placed at the point corresponding to the top of the barrier and all
particles go down the side where the quantity $q$ grows indefinitely.

Given the current $j$ the stationary distribution function $w(x)$ is
defined by Eq.~(\ref{37}), according to which, $w(x)\approx w_{0}(x)$
in the subcritical region $q<q_{c}$, while in the supercritical range
$q\gg q_{c}$ we have $w_{0}(x)\gg w(x)$ due to indefinite increase of
$w_{0}(x)$. As far as the stationary distribution is concerned, it
can be derived from the current definition (\ref{35}), where the last
diffusion term is negligible for supercritical heats:
$j\approx -(\partial V/\partial x)w$. The result is that
the probability $w(q)$ remains almost unaltered, $w(q)\approx
w(q_{c})$, in the range from $q_{c}$ up to a boundary value $q_{g}$
and $w(q)\approx 0$ at $q>q_{g}$ \cite{17}. The growth of $q_{g}$ is governed
by the equation \begin{equation}
\frac{{\rm d}q_{g}}{{\rm d}\kappa}= \chi\frac{q_{g}-q_{c}}{q_{g}^{2}}.
\label{40} \end{equation}

Since the above picture is essentially statistical,
it enables the critical heat effect $q_{c}$ for the transition
point to be found. Indeed, when the definition of the macroscopic current
$J$ in Eq.~(\ref{25a}) is compared to that of the microscopic current
$j$ in Eq.~(\ref{35}), it is apparent that they differ from one
another only by the factor $N^{(a-1)/a}\equiv s_{0}^{a-1}$ dependent
on the total number of embryos $N$ (see Eq.~(\ref{22a})). On this
basis, the last expression of Eq.~(\ref{31}) and Eq.~(\ref{38}) at
$G=0,\,\chi\ll 1$ give the desired result:
\begin{equation}
q_{c}=Q\exp{\left(
-\frac{(1-\beta)^{-1}}{2\chi}\right)},
\label{41}
\end{equation}
where the pre-exponent factor $Q$ determines the probability of the
barrier penetrating and cannot be calculated within the framework of
the presented approach. Eq.~(\ref{41}) predicts the slow growth of
the critical thermal effect $q_{c}$ of embryo with the thermometric
conductivity coefficient $\chi$.

\section{Time dependence of crystal formation probability}\label{sec5}

Since the ensemble of hierarchically subordinated crystallization
centers represents a self-similar set, the probability distribution
of embryos $P(Q,s)$ in the course of SOC process is a homogeneous
function of $s$ \cite{10}: \begin{equation} P(Q,s)=s^{-\tau}w(q),
\label{42} \end{equation} where $w(q)$ is the stationary
distribution of embryos and $\tau$ is the positive exponent. Physically,
Eq.~(\ref{42}) implies that the heat effect $Q$, being measured by
the scale $N^{b/a}$, equals the heat effect of an embryo
formation $q$ in accordance with Eq.~(\ref{28}).

In this section we are aimed to define the probability of hierarchical
crystallization leading to the formation of a fractal cluster (see
Fig.~1). As it has been clarified in Sec.~\ref{sec4}, this process can
be conceived of as diffusion in ultrametric space that makes the
distribution (\ref{42}) mounted. In order to find the conditional
probability $\overline{{\cal P}}(t)$ that no crystallization
will begin at time $t$ one has to integrate over $s$ the distribution
(\ref{42}) weighted with the function \begin{equation}
p_{s}(t)=\exp{(-t/t(s))},
\qquad t(s)=t_{0}\exp{[Q(s)/\chi]} \label{43} \end{equation}
descriptive of Debye relaxation with the time $t(s)$ governed by the
barrier height $Q(s)$ (see Eq.~(\ref{27a}))
and a microscopic time $t_0$ determined below.  By using the steepest
descent method, it is not difficult to derive the late time ($t\gg
t_{ef}$) asymptotic formula \begin{equation} \overline{{\cal
P}}(t)=\left(\frac{q}{\chi}\right)^{\tau/b}
\left[1-\left(\frac{\chi}{q} \ln{\frac{t}{t_{ef}}} \right)^{-1/b}
\right]^{-\tau},\qquad t_{ef}\equiv\frac{\tau}{b}\left(
\frac{q}{\chi} \right)^{1/b}t_{0}. \label{44} \end{equation}
Eq.~(\ref{44}) has been obtained by assuming that the condition $1\ll
s_{m}\le s_{0}$ is met, where $s_{m}$ denotes the location of the
maximum of integrand and obeys the equation \begin{equation}
\frac{\chi\tau}{bq}\frac{(1-x)^{1+b}}{x}= \frac{t}{t_{0}}\exp{\left(
-\frac{q}{\chi}(1-x)^{-b} \right)},\qquad x\equiv
\frac{s_{m}}{s_{0}}. \label{45} \end{equation} Taking into
consideration the scaling relation for the number of hierarchical
levels $s_0$, which is the cut-off parameter \cite{10}
\begin{equation}
s_{0}\propto \left(q_{c}-q\right)^{-1/\sigma},\qquad \sigma>0
\label{46} \end{equation} we readily come to the conclusion that the
condition is satisfied provided \begin{equation} q-q_{c}\ll q,\qquad
t\gg t_{ef} \exp{\left(\left(q_{c}/\chi\right)^{-1/b}-1
\right)^{-b}}. \label{47} \end{equation} Clearly, from Eq.~(\ref{47})
the heat effect $q$ in Eqs.~(\ref{44}), (\ref{45}) can be replaced by
$q_{c}$.  Note that in accord with Eq.~(\ref{44}) the crystallization
probability ${\cal P}(t)\equiv 1-\overline{\cal P}(t)$
logarithmically increases in time up to the value ${\cal
P}=1-(q_{c}/\chi)^{\tau/b}$. The condition ${\cal P} \ge 0$ is
satisfied if in Eq.~(\ref{41}) factor $Q=({\rm e}/2)(1-\beta)^{-1}$
and thermometric conductivity coefficient is restricted by the
maximum value $\chi_{0}=(1/2)(1-\beta)^{-1}$.

\section{Discussion}\label{sec6}

In accordance with above approach, given in Sec.~\ref{sec3}, the
transition from the cold crystallization mode of amorphous material
to the explosive crystallization mechanism represents the
self-organization process realized as a first-order transition. The
crystallization front velocity $u$ represents the order parameter,
the temperatures difference $T$ at the crystallization front and
thermostat acts as the conjugate field, and the difference $f$ of
specific thermodynamic potentials  of amorphous and crystalline
states is the control parameter. Equations (\ref{14})--(\ref{16})
are derived assuming the degrees of freedom $u$, $T$ and $f$ to be
dissipative. In addition, the positive feedback between $u$ and $f$
is taken into consideration as the reason behind the
self-organization, whereas the negative feedback between $u$ and $T$
is a manifestation of Le Chatelier principle.  The system is driven
by the parameter $\theta$, whose value represents the thermal energy
of crystallization (or externally inputted energy). When
$\theta$ is above the critical value (\ref{20}), the effective
potential (\ref{19}) assumes additional minimum value $-q<0$ at
point $u=u_e$ and maximum one $U$ at $u=u^m$ (stationary values of
crystallization front velocity $u$ and thermodynamic factor $f$ are
given by Eqs.~(\ref{22}), (\ref{24})). The explosive crystallization
process is preferable in potential provided the minimum effective
potential becomes negative ($q>0$). The height of the energy barrier
$U$ of the effective potential (\ref{19}) defines characteristic time
in the last Eq.~(\ref{43}) for hot crystallization center to be
formed \begin{equation} t_{0}\approx\tau_{D}\exp{(U/\Delta)},
\label{48} \end{equation} where $\Delta$ is the variance of $\theta$
and $\tau_D\sim 10^{-12}$s is the Debye time.

The nucleating crystals form statistical ensemble of
hierarchically subordinated objects, characterized by heat
$q$ and distances $s$ in ultrametric space (crystalline cluster size
\cite{10}).  Since the crystallization represents the
effective diffusion over hierarchical tree nodes, then, similar to
Brownian particle with coordinate $q$ at time $s$, the ensemble can
be described by Langevin equation (\ref{33}) subjected to the noise
Eq.~(\ref{34}) with $\chi$ being the effective diffusion coefficient
(thermometric conductivity) and corresponding Fokker-Planck
equation (\ref{35}).  The stationary heat distribution and the
steady-state current are given by Eqs.~(\ref{37}), (\ref{38}). The
condition of current conservation Eq.~(\ref{25a}) yields the heat
distribution (\ref{27a}) in the ultrametric
space. The ensemble of embryos, being weakly dependent
on $s$, is governed by the effective potential (\ref{32}) that
reaches its maximum at the critical heat (\ref{41}) (see
Fig.~4). So, the explosive crystallization requires supercritical
heat effect, $q>q_{c}$, to surmount the
barrier $V_{0}$ with the characteristic time (cf. Eq.~(\ref{39}))
\begin{equation} T\approx t_{0}\exp{(V_{0}/\chi)}. \label{49}
\end{equation} This picture bears some resemblance with the formation
process of supercritical embryo in the theory of a first-order
phase transitions \cite{16}. In the course of phase transformation
the next stage is the growth of the embryo due to the diffusion
increase of the heat effect $Q(s)$ in ultrametric space. As a result,
we have the logarithmically slow large time asymptotic for the
probability of the total cluster formation: \begin{equation} {\cal
P}(t)=1-\overline{\cal P} \left[1-\overline{\cal P}^{1/\tau}\left(
\ln{\frac{t-T}{t_{ef}}} \right)^{-1/b} \right]^{-\tau},\qquad
t_{ef}\equiv(\tau/b) \overline{\cal P}^{1/\tau} t_{0}, \label{50}
\end{equation} where time $t$ is counted from the instant $T$,
Eq.~(\ref{49}), and $\overline{\cal P}$ is the minimum probability
that no crystallization will occur:
\begin{equation} \overline{\cal
P}=\left( \frac{\chi_{0}}{\chi} \right)^{\tau/b}\exp{\left[
-\frac{\tau}{b}\left( \frac{\chi_{0}}{\chi}-1 \right) \right]}.
\label{51} \end{equation}
From Eq.~(\ref{51}) the probability is determined by the ratio of the
noise intensity $\chi$ (see Eq.~(\ref{34})) and its maximum value
$\chi_{0}=(1/2)(1-\beta)^{-1}$. The key point is that the maximum
probability ${\cal P}\equiv 1- \overline{\cal P}$ of crystallization
is completely suppressed under great thermometric conductivity
coefficient $\chi$ (Fig.~5).

\newpage

\begin{center}
{\bf FIGURE CAPTIONS}
\end{center}

\begin{description}

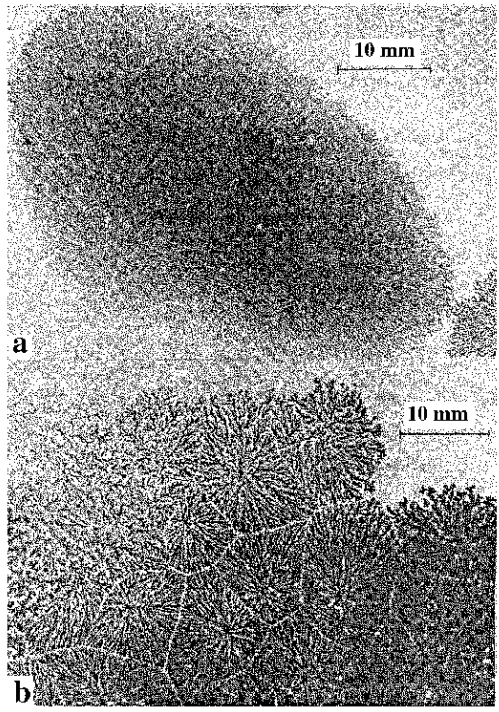
\item[Fig.~1.] The patterns which arise in the layer  of  amorphous
powder  of  Ge  at  the explosive crystallization: (a) from single
center; (b) from several centers which arise spontaneously.

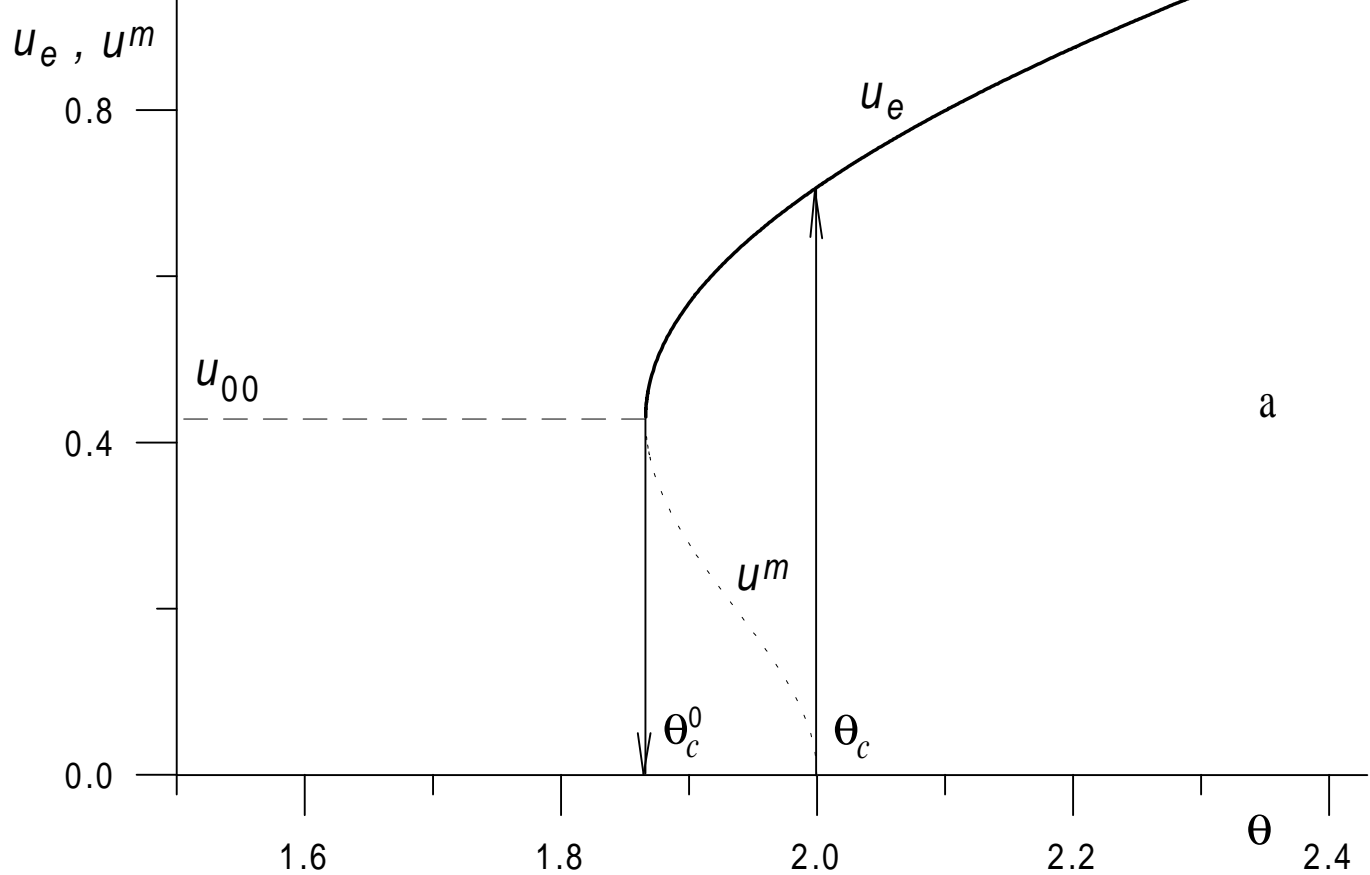
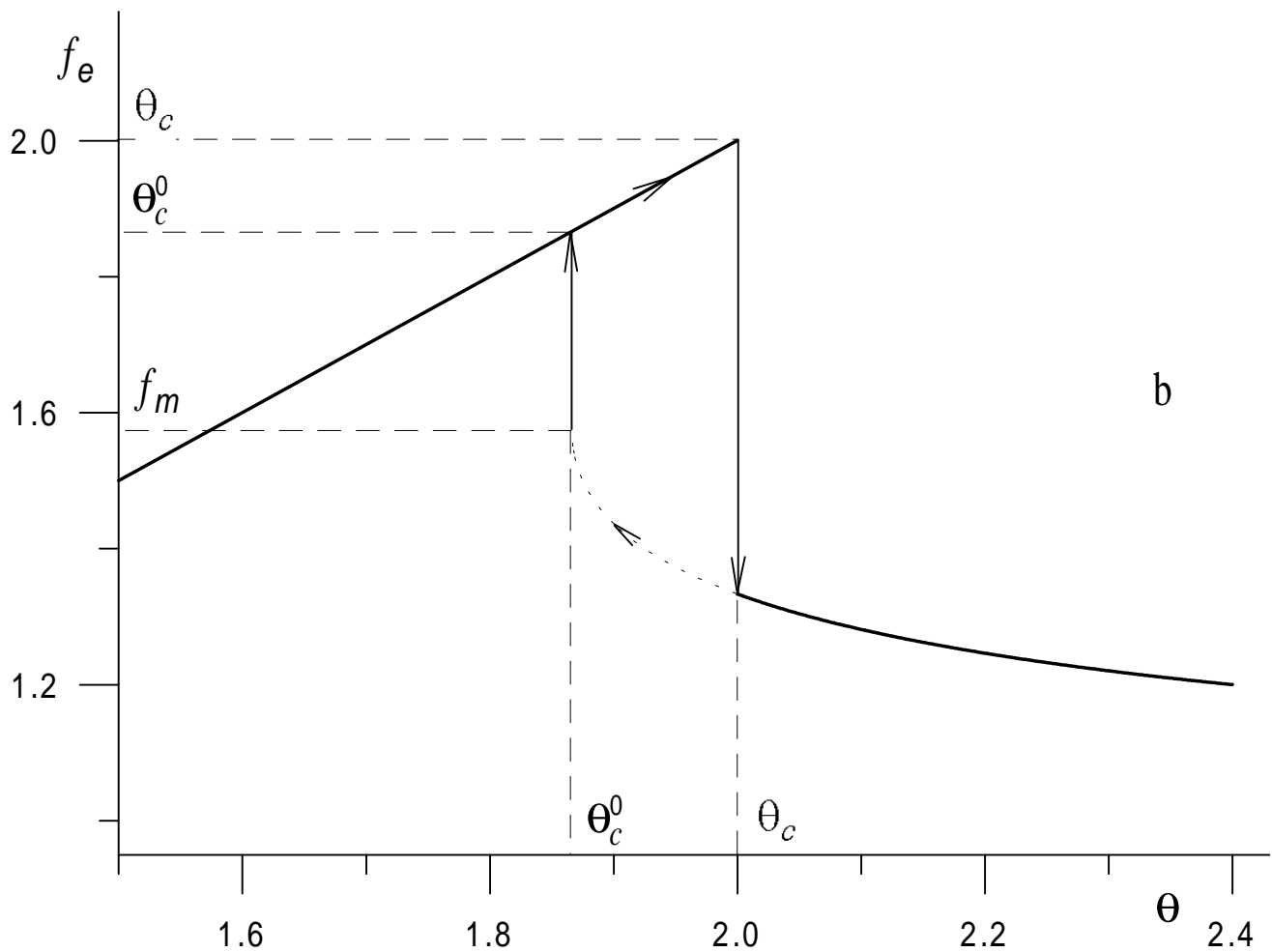
\item[Fig.~ 2.] (a) The dependence of the steady-state velocities of
crystallization front  on  the transition heat (the solid curve
corresponds to the stable state $u_e$, the dotted curve  meets the
unstable one $u^m$). (b) The dependence of the stationary difference
$f_e$ of specific thermodynamic phase potentials on the transition
heat. The arrows indicate the hysteresis loop.

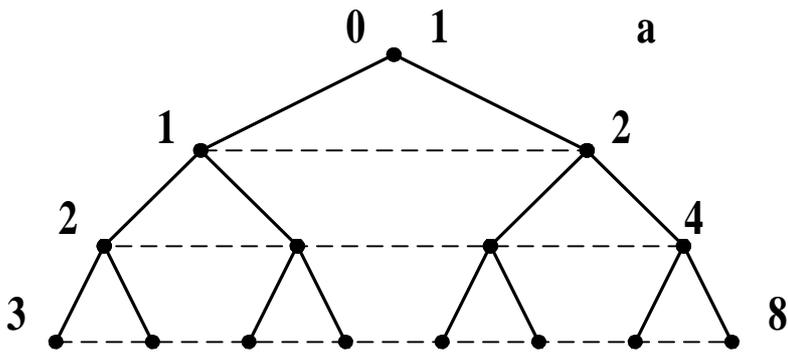
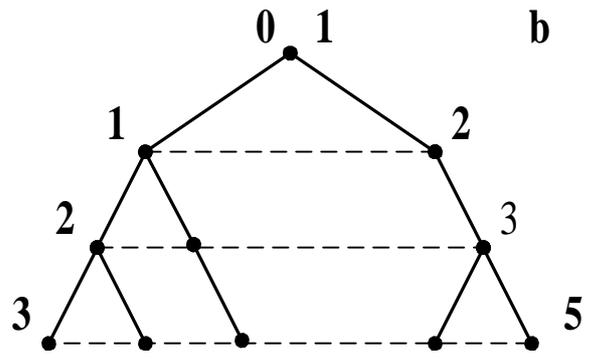
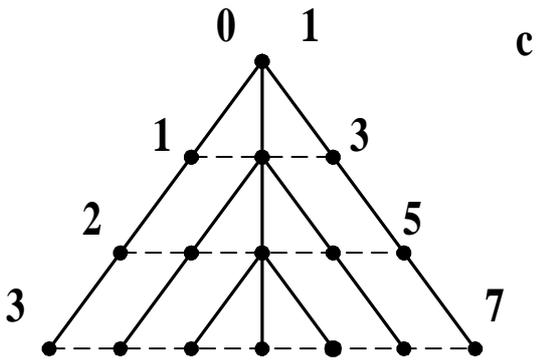
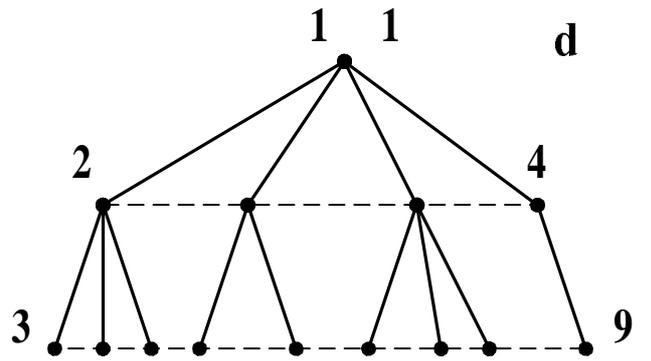
\item[Fig.~ 3.] Different types of hierarchical trees
(the level number is indicated at left, corresponding number of
nodes -- at right): a) regular tree with $j=2$; b) Fibbonachi tree;
c) degenerate tree with $j=3$; d) irregular tree.

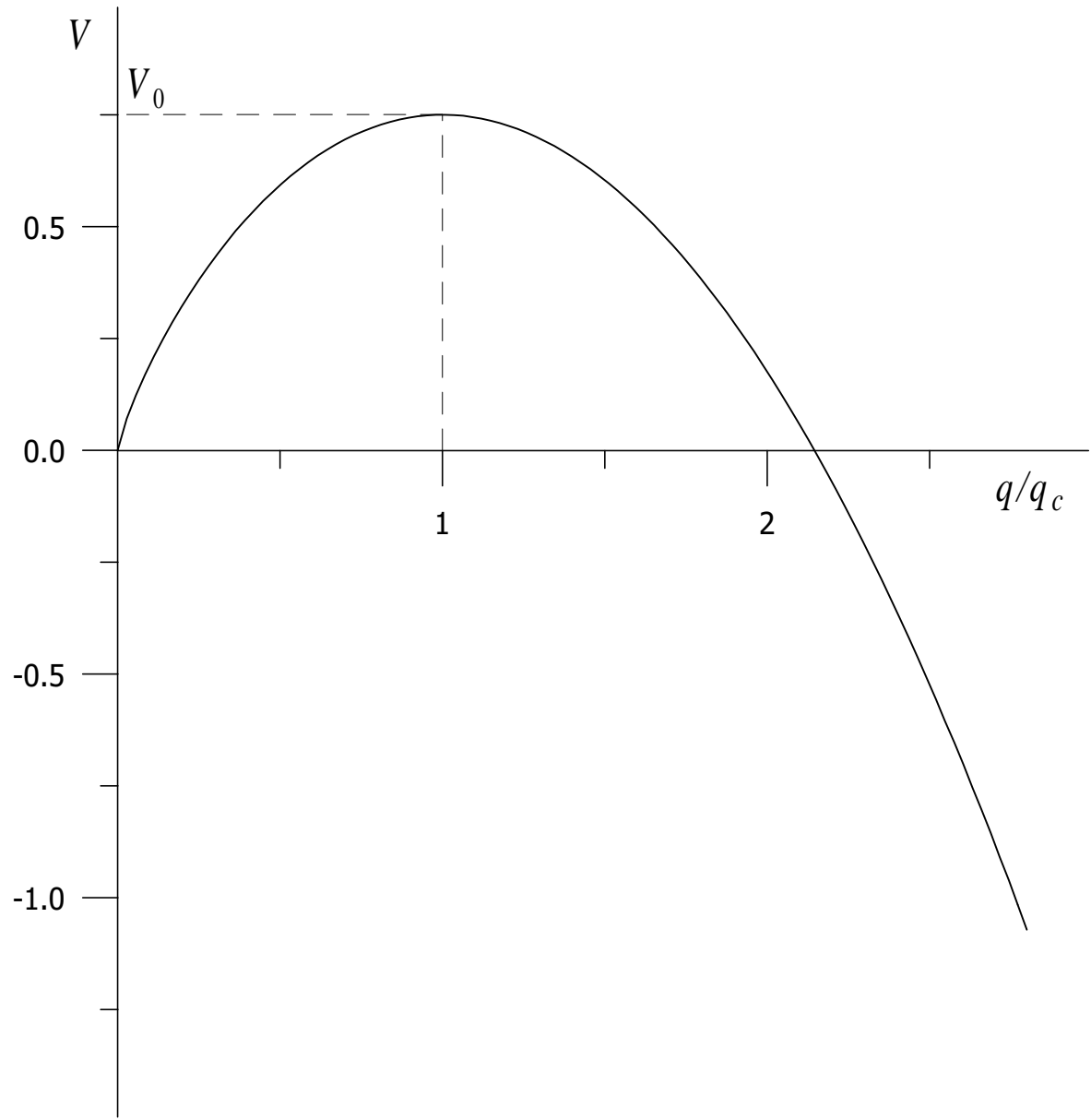
\item[Fig.~4.] The form of the effective potential (\ref{32})
at $\beta=0.2$.

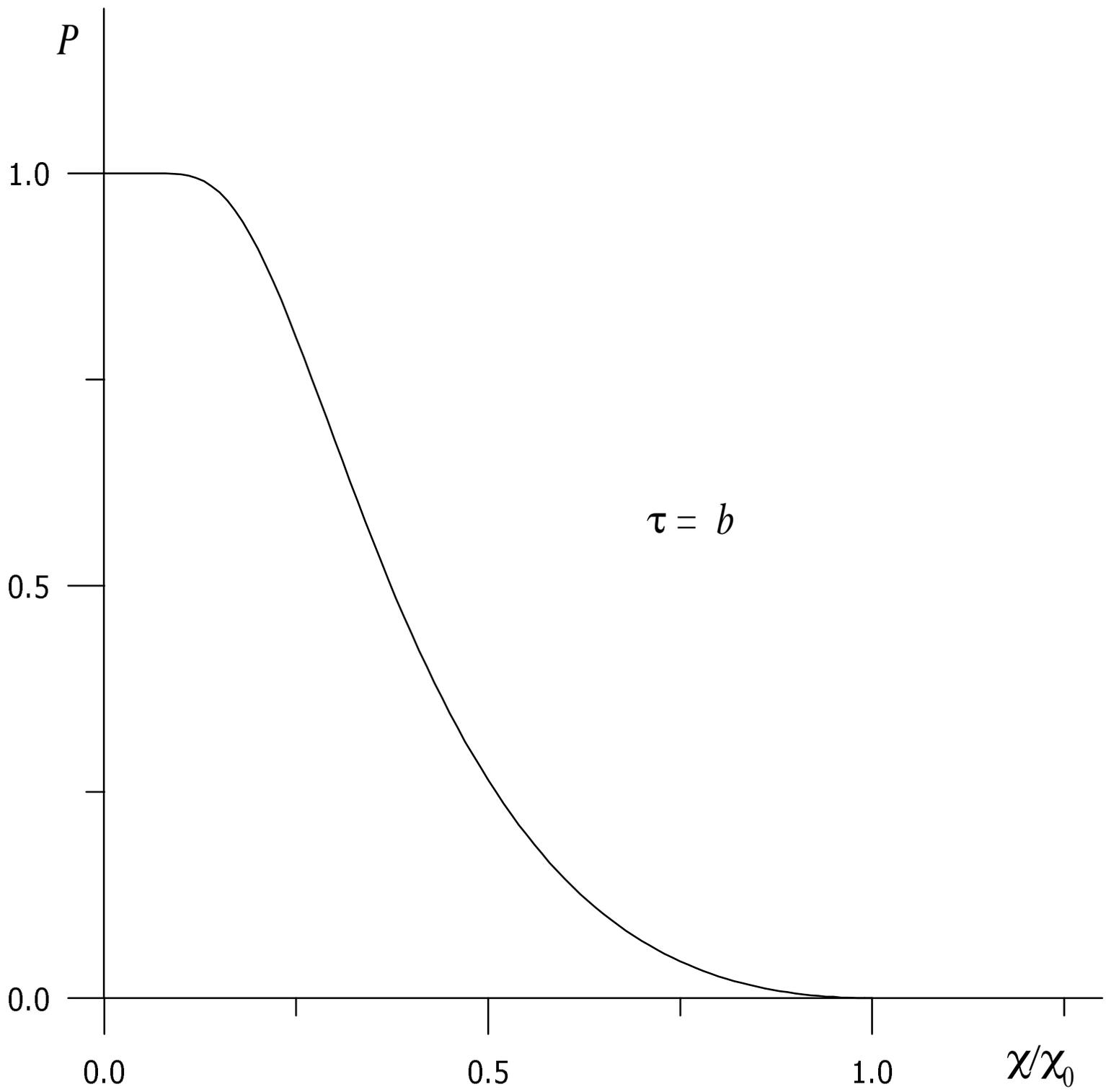
\item[Fig.~5.] The dependence of the maximum probability ${\cal P}$
of crystallization on the thermometric conductivity $\chi$.

\end{description}
\end{document}